\documentclass{entcs}
\usepackage{amsmath}
\usepackage{amssymb}
\usepackage{epsfig}
\usepackage{graphicx}
\usepackage{epstopdf}
\def\slashchar#1{\setbox0=\hbox{$#1$}     		
   \dimen0=\wd0                                 	
   \setbox1=\hbox{/} \dimen1=\wd1               	
   \ifdim\dimen0>\dimen1                        	
      \rlap{\hbox to \dimen0{\hfil/\hfil}}      	
      #1                                        	
   \else                                        	
      \rlap{\hbox to \dimen1{\hfil$#1$\hfil}}   	
      /                                         	
   \fi}      

\begin{document}

\renewcommand{\vec}{\boldsymbol }
\newcommand{\rap}{Y}
\newcommand{\qw}{Q_\mathrm{w}}
\newcommand{\nw}{n_\mathrm{w}}
\newcommand{\qp}{Q_\mathrm{p}}
\newcommand{\nt}{N_\mathrm{t}}
\newcommand{\na}{N_\mathrm{a}}
\newcommand{\nb}{N_\mathrm{b}}
\newcommand{\dt}{\Delta_\mathrm{t}}

\begin{frontmatter}

\title{The effects of topological charge change in 
heavy ion collisions:\\ ``Event by event $\mathcal{P}$
and $\mathcal{CP}$ violation''}

\author[BNL]{Dmitri E. Kharzeev, } 
\author[BNL,RBRC]{Larry D. McLerran}
and 
\author[BNL]{Harmen J. Warringa}

 \address[BNL]{Department of
  Physics, Brookhaven National Laboratory, Upton NY 11973, USA
 }
\address[RBRC]{RIKEN BNL Research Center,
 Brookhaven National Laboratory, Upton, NY 11973, USA}

\date{\today}

\begin{abstract}
  Quantum chromodynamics (QCD) contains field configurations which can
  be characterized by a topological invariant, the winding number
  $\qw$. Configurations with nonzero $\qw$ break the charge-parity
  ($\mathcal{CP}$) symmetry of QCD.  We consider a novel mechanism by which these
  configurations can separate charge in the presence of a background
  magnetic field -- the "Chiral Magnetic Effect".  We
  argue that sufficiently large magnetic fields are created in heavy
  ion collisions so that the Chiral Magnetic Effect causes
  preferential emission of charged particles along the direction of
  angular momentum.  Since separation of charge is $\mathcal{CP}$-odd,
  any observation of the Chiral Magnetic Effect could provide a clear
  demonstration of the topological nature of the QCD vacuum.  We give
  an estimate of the effect and conclude that it might be observed
  experimentally.
\end{abstract}

\end{frontmatter}

\section{Introduction}

A remarkable property of gauge theories is the existence of
topologically non-trivial configurations of gauge fields
\cite{Belavin1975}. In Minkowski space-time, these configurations are
associated with tunneling between different states which are related
by topologically non-trivial gauge transformations 
\cite{thooft1976,rebbi1976}.  These different states are
characterized by the Chern-Simons topological charge \cite{chern1974}.
Transitions between these different states which involve changes in
the topological charge induce anomalous processes, that is, processes
which are forbidden by the classical equations of motion of the
theory, but when the theory is quantized, become allowed
\cite{adler1969,jackiw1969}. The amplitudes of such transitions
vanish order by order in perturbation theory, and are associated with
intrinsically non-perturbative phenomena involving many quanta. In
electroweak theory, such transitions are associated with $B+L$
violation, where $B$ is baryon number and $L$ is lepton number.  In QCD,
they are tied to chirality non-conservation.  Chirality conservation
is an exact symmetry of the classical equations of motion of QCD and
of perturbation theory in the zero quark mass limit.  The zero quark
mass limit should be a good approximation for high energy processes

It was originally thought that non-perturbative phenomena associated
with the underlying topology were suppressed by factors such as
$e^{-2\pi/\alpha}$ where $\alpha$ is the interaction strength of the
gauge theory.  This is typical of results from computation of quantum
tunneling between states with different topological charge.  It was a
surprise when Manton and Klinkhamer proposed that for processes at
high temperature, that the exponential suppression might disappear
\cite{manton1983,Klinkhamer1984}.  This is because at high
temperatures, there is sufficient energy to pass classically over the
barrier which separates states of different topological charge.
Kuzmin, Rubakov and Shaposhnikov argued that the rate would be
sufficiently large so that these processes might be important for
electroweak baryogenesis \cite{Kuzmin1985,Shaposhnikov1987}.
The rates for such processes were computed \cite{Arnold1987}, and it
was eventually understood that such rates were not exponentially
suppressed because they involved the coherent interaction of many
low--energy quanta in an initial state going to many quanta in a final
state \cite{Arnold1988}.  Such rates should be large therefore not
only at finite temperature but also in out--of--equilibrium processes
which involve many particles.  It was also shown that the computations
for electroweak baryon number violation have a counterpart in helicity
non-conservation in QCD \cite{McLerran1990}.

\bigskip

An interesting feature of the transitions changing topological
charge is that they involve ${\cal P}$-- and ${\cal C P}$--odd field
configurations.  Because of this, ${\cal P}$-- and ${\cal C
  P}$--invariance of strong interactions is no longer "natural".
These symmetries can however be imposed dynamically by postulating the
existence of axions \cite{axion,KSVZ,DFSZ}; see \cite{review} for
reviews. In any case, ${\cal P}$-- and ${\cal C P}$--invariance of
strong interactions has to be related to the vacuum structure. It is
thus possible that an excited vacuum domain which may be produced in
heavy ion collisions as proposed by Lee and Wick \cite{LeeWick} can
break parity and and ${\cal C P}$ spontaneously
\cite{Lee:1973iz}. Some consequences of such a scenario had been
studied by Morley and Schmidt \cite{Morley:1983wr}, who proposed as a
signature ${\cal C P}$--odd correlations between the spins of the
protons produced in a heavy ion collision. Other realizations of an
excited vacuum domain (including the ones with parity--odd
fluctuations) have been considered in Refs \cite{DCC}.

\bigskip 

It was proposed that in the vicinity of the deconfinement phase
transition QCD vacuum can possess metastable domains leading to ${\cal
  P}$ and ${\cal CP}$ violation \cite{Kharzeev1998}.  It was also
suggested that this phenomenon would manifest itself in specific
correlations of pion momenta \cite{Kharzeev1998,KP}.  (For related
studies of metastable vacuum states, especially in supersymmetric
theories, see \cite{SUSY1,SUSY2,SUSY3}).  The existence of ${\cal
  P}$--odd bubbles had been inferred from the analysis of an effective
chiral theory incorporating axial anomaly \cite{Kharzeev1998}; such
bubbles may also be viewed as the space--time domains occupied by
gauge field configurations with non--trivial winding number.

The existence of metastable ${\cal P}$--odd bubbles does not
contradict the Vafa--Witten theorem \cite{VW} stating that ${\cal P}$
and ${\cal CP}$ cannot be broken in the true ground state of QCD for
$\theta =0$. Moreover, this theorem does not apply to QCD matter at
finite isospin density \cite{SonMisha} and finite temperature
\cite{Cohen}, where Lorentz--non-invariant ${\cal P}$--odd operators
are allowed to have non-zero expectation values. For matter at zero
temperature and finite baryon density, Migdal \cite{Migdal} long time
ago advocated the possibility of parity--breaking pion condensate
phase (for a recent work on $\mathcal{P}$--violation in cold dense matter, see
\cite{Andrianov:2007kz} and references therein). Degenerate vacuum
states with opposite parity were found \cite{PR} in the
superconducting phase of QCD. A parity broken phase also exists in
lattice QCD with Wilson fermions \cite{Aoki}, but this phenomenon has
been recognized as a lattice artifact for the case of mass--degenerate
quarks; spontaneous ${\cal P}$ and ${\cal CP}$ breaking similar to the
Dashen's phenomenon \cite{Dashen} can however occur for non--physical
values of quark masses \cite{Creutz}. ${\cal P}$--even, but ${\cal
  C}$--odd metastable states have also been argued to exist in hot
gauge theories \cite{Chris}. 
The conditions for the applicability of
Vafa-Witten theorem have been repeatedly re--examined in recent years
\cite{VWth}.

Several dynamical scenarios for the decay of ${\cal P}$--odd bubbles
have been considered \cite{decay}, and numerical lattice calculations
of the fluctuations of topological charge in classical Yang--Mills
fields have been performed \cite{Raju,Tuomas}. A closely related topic
is the role of topological effects in high--energy scattering and
multi--particle production which has been studied in Refs
\cite{Moch:1996bs,Kharzeev:2000ef,Shuryak:2000df,Nowak:2000de,Janik:2002nk}.
The studies of ${\cal P}$-- and ${\cal CP}$--odd correlations of pion
momenta \cite{Voloshin,Sandweiss} have shown that such measurements
are in principle feasible but would require large event samples.

\vskip0.3cm

Some time ago, it was pointed out that the presence of non--zero
angular momentum (or equivalently of magnetic field) in heavy ion
collisions makes it possible for ${\cal P}$-- and ${\cal CP}$--odd
domains to induce charge separation in the produced particles
\cite{Kharzeev2006}. This is because the ${\cal P}$-- and ${\cal
  CP}$--breaking term carries net chirality and thus generates
asymmetry between left-- and right--handed fermions $\sim {\bar
  \psi}_L \psi_R - {\bar \psi}_R \psi_L \neq 0$. In the presence of
a magnetic field a ${\cal P}$-- and ${\cal CP}$--odd domain can generate
chirality not by flipping the spins of the quarks, but by inducing
up--down asymmetry (as measured with respect to the symmetry axis of
the angular momentum) in the production of quarks and anti--quarks
\cite{Kharzeev2006}. The magnitude of the effect has been estimated,
and found to be within experimental reach.  A very useful
experimental observable measuring the charge separation with respect
to reaction plane (which is perpendicular to the angular momentum and
magnetic field axis) has been proposed by Voloshin
\cite{Voloshin:2004vk} shortly afterwards. The first result of a
dedicated experimental study of the effect has been recently presented
by the STAR Collaboration at RHIC \cite{Selyuzhenkov:2005xa}.

In Ref. \cite{Kharzeev:2007tn} it has been noticed that the
 effect of charge separation can also be understood in the following
way. At finite $\theta$--angle (as present effectively in a ${\cal
  P}$-- and ${\cal CP}$-odd vacuum domain), a magnetic field ${\vec
  B}$ induces through the axial anomaly a parallel electric field
$\sim \theta\ {\vec E}$, and a corresponding electric dipole moment. 
One may consider as a theoretical
illustration of this phenomenon the transformation, at finite
$\theta$, of a magnetic monopole into a "dyon" with an electric charge
$\sim - e\ \theta/2 \pi$ demonstrated by Witten
\cite{Witten:1979ey}. In an off-central heavy ion collision, a ${\cal
  P}$-- and ${\cal CP}$-odd vacuum domain was shown
\cite{Kharzeev:2007tn} to induce an electric field perpendicular to
the reaction plane of magnitude $E_z \sim - (e\ \theta / 2 \pi)\ l$,
where $l \sim b$ is the angular momentum in a collision at impact
parameter $b$. The corresponding electric charge separation has been
estimated and found amenable to experimental observation.

\bigskip

In this paper we discuss a novel mechanism for charge separation. The
topological charge changing transitions provide the ${\cal P}$-- and
${\cal CP}$-- violation necessary for charge separation.  The variance
of the net topological charge change is proportional to the total
number of topological charge changing transitions. Hence if
sufficiently hot matter is created in heavy ion collisions so that
topological charge transitions can take place, we expect on average in
each event a finite amount of topological charge change.

Charge separation needs a symmetry axis along which the separation can
take place. The only symmetry axis in a heavy ion collision is angular
momentum which points in the direction perpendicular to the reaction
plane. In central collisions there is no symmetry axis, so in that
case charge separation should vanish.

If charge separation would manifest itself mainly into pions it should
violate isospin symmetry (interchange of up and down quarks). This is
because under an isospin transformation a $\pi^+$ becomes a $\pi^-$
meson, so in an isospin symmetric situation $\pi^+$ should behave
similar to $\pi^-$, hence in that case they can never
separate. Isospin violation can occur because up and down quarks have
a different mass, because our initial state could be isospin
asymmetric, and because up and down quarks have different charge, or
because of the presence of an electromagnetic field.

Since QCD contains configurations that break the
$\mathcal{CP}$ symmetry, but not the $\mathcal{C}$ symmetry, the charge
separation mechanism we are looking for should not violate $\mathcal{C}$.
However, a particle with positive charge transforms to  a particle
with negative charge under charge conjugation.  In order to maintain
$\mathcal{C}$ invariance, some important ingredient of the mechanism
that separates charge should transform under $\mathcal{C}$ as
well. The only possibility lies in the initial state, since the
incoming ions have a positive electric charge, they transform under
$\mathcal{C}$. So our mechanism should somehow depend on the electric
charges of the ions.

Summarizing, the charge separation mechanism should provide a symmetry
axis, it should violate isospin and depend on the charges of the
colliding ions. Therefore it is very natural that the
(electromagnetic) magnetic field is crucial for charge separation.
This magnetic field depends on the electric charges of the colliding
nuclei, so that $\mathcal{C}$ invariance is maintained. The magnetic
field provides a symmetry axis; midway between the two colliding
nuclei, the magnetic field is parallel to the angular momentum
vector. Moreover the magnetic field causes violation of isospin since
up and down quarks will be polarized oppositely in a magnetic field.

We can understand the effect we will discuss in this article in a
simple example.  Suppose we have a box with some net topological
charge.  This means that due to the anomaly, there is a net helicity
for the quarks inside the box.  If the box is at high enough
temperature, then perturbative interactions which involve spin flip
are negligible, since $T \gg m_{\mathrm{quark}}$, so the net helicity
is directly proportional to the net topological charge.  Now imagine
we apply an electrodynamic magnetic field.  This will orient the spin
of positively charged quarks (and anitquarks) along the direction of
the magnetic field, due to the magnetic moment interaction. Likewise,
the spins of negatively charged quarks (and anitquarks) will become
oriented opposite to the direction of magnetic field.  Now because
there is net helicity, the orientation of spin is preferentially in
the direction of momentum of the particles.  Positively charged quarks
and antiquarks move in the direction of the magnetic field, and
negatively charged quarks and antiquarks move opposite to it.
Therefore an electromagnetic current is set up in the direction of the
magnetic field.  We therefore call this the Chiral Magnetic
Effect. The Chiral Magnetic Effect induces a vector current. For
systems of a finite size, this current results in charge separation or
an induced electric dipole moment.

The separation is as required $\mathcal{C}$ even, since under
$\mathcal{C}$ both the magnetic field and the vector current change
sign.  Under $\mathcal{P}$, the current changes sign, but the magnetic
field does not, so it is $\mathcal{P}$ odd, and $\mathcal{CP}$ odd.

This article is organized as follows. In Section 2 and 3 we review the
axial anomaly and the QCD vacuum structure respectively. In Section 4
we discuss in detail the Chiral Magnetic Effect. Section 5 is devoted
to the consequences of this effect in heavy ion collisions. In Section
6 we define some observables and Section 7 contains a discussion of
our results.  The derivation of the magnetic field is discussed in
Appendix~A.  An alternative derivation of the Chiral Magnetic Effect
is given in Appendix~B.

\section{The Axial Anomaly}

In this section, we review some properties of topologically
non-trivial gauge fields.  We also review the relationship between
topology and anomalous processes.

All gauge field configurations which have finite action
can be characterized by a topological invariant, the winding number
$\qw$. The winding number is an integer and given by
\begin{equation}
\qw
= \frac{g^2}{32\pi^2} \int \mathrm{d}^4 x\, 
F_{\mu \nu}^a \tilde F^{\mu\nu}_a
\in \mathbb{Z}.
\end{equation}
Here $g$ denotes the QCD coupling constant 
with generators normalized as $\mathrm{tr}\, t_a t_b = \delta_{ab} / 2$.
The gluonic field tensor and its dual are respectively
$F_{\mu\nu}^a$ and $\tilde F^a_{\mu\nu} = 
\tfrac{1}{2} \epsilon_{\mu\nu}^{\phantom{\mu\nu}\rho\sigma} F^a_{\rho \sigma}$.
Configurations with nonzero $\qw$ lead to non-conservation of axial
currents even in the chiral limit. This can be seen from the axial
Ward-identity which reads
\begin{equation}
 \partial^\mu j_\mu^5 = 2 \sum_f m_f \langle \bar \psi_f 
i \gamma_5 \psi_f \rangle_A 
 - \frac{N_f g^2}{16\pi^2} F_{\mu\nu}^a \tilde F^{\mu\nu}_a,
\end{equation}
here $N_f$ denotes the number of quark flavors, $\psi_f$ a quark field,
and $m_f$ the mass of a quark. 
The axial current $j_\mu^5$ is given by
\begin{equation}
 j_\mu^5 = \sum_f \langle \bar \psi_f \gamma_\mu \gamma_5 \psi_f \rangle_A.
\end{equation}
We write $\langle O \rangle_A$ to denote an average over fermionic
fields only in the background of a particular gauge field
configuration $A_\mu$. To obtain the full expectation value $\langle O
\rangle$ one still has to perform the integration over the gluonic
fields $A_\mu$. In this article we will only consider the chiral limit,
i.e.~$m_f = 0$. 

Assuming that initially at $t=-\infty$ we have an equal number of
right-handed and left-handed fermions\footnote{With fermions we mean
quarks and antiquarks. We write $N_R$ for the
total number of right-handed fermions, which is the sum of the number
of right-handed particles and right-handed antiparticles. 
Here right-handed massless fermions have spin and momentum 
parallel, while left-handed massless fermions have spin and momentum
antiparallel. 
The volume
integral over $\psi_R^\dagger \psi_R$ however, is equal to the total
number of right-handed particles minus the total number of left-handed
antiparticles.}, i.e.\ $N_R = N_L$, it follows from the axial Ward
identity that at $t=\infty$
\begin{equation}
(N_L - N_R)_{t = \infty}  
= 2 N_f \qw.
\end{equation}
This shows that configurations with positive $\qw$ convert
right-handed fermions into left-handed ones.

Let us now write the number of right-
or left-handed fermions of a particular flavor as $N_{R,L}^f$. So
$N_{R,L} = \sum_f N_{R,L}^f$.  Assuming that initially for each flavor
separately $N_R^f = N_L^f$, one can show that $t = \infty$
\begin{equation}
(N^u_L - N^u_R)_{t = \infty}  = (N^d_L - N^d_R)_{t = \infty} 
 \label{eq:quarkindep}
\end{equation}
where in this equation $u$ and $d$ can denote any quark flavor. This
implies that in the chiral limit, the number of converted up
and down quarks is equal.

\section{The QCD Vacuum}
In the classical vacuum of QCD the gauge field has to be a pure gauge
so that the energy density is minimal.   
In the temporal gauge ($A_0 = 0$) this implies $A_i(\vec
x) = \tfrac{i}{g} U(\vec x) \partial_i U^{\dagger}(\vec x)$, where
$U(\vec x)$ is an element of the gauge group $\mathrm{SU}(3)$.  The
different classical vacua can be characterized by a topological
invariant, the winding number $\nw$ which is an integer and given by
\begin{equation}
 \nw = \frac{1}{24\pi^2}
\int \mathrm{d^3}x 
\,
\epsilon^{ijk} \mathrm{tr}
\left[
(U^\dagger \partial_i U)
(U^\dagger \partial_j U)
(U^\dagger \partial_k U)
\right].
\end{equation}
Now one can show that if a gauge field configuration goes to a pure
gauge at infinity and has nonzero $\qw$ it induces a transition from
one classical vacuum to another, more precisely
\begin{equation}
\qw = \nw(t=\infty) - \nw(t=-\infty).
\end{equation}

At zero temperature such transition requires tunneling through a
potential barrier, which will suppress the transition rate
exponentially. In this case the main contribution to the transition
rate comes from fluctuations around instantons (which are minima of
the classical Euclidean action). It was
found by 't Hooft \cite{thooft1976} (see also \cite{Bernard1979}) in
QCD without fermions that the number of transitions $\nt^\pm$ 
with $\qw= \pm 1$ per unit volume and time equals
\begin{equation}
\frac{ \mathrm{d} \nt^\pm}
{\mathrm{d}^3 x \,\mathrm{d}t \, \mathrm{d}\rho} 
= 
0.0015 \left(\frac{2\pi}{\alpha_S} \right)^6
\exp \left( - \frac{2\pi}{\alpha_S} \right)
\frac{1}{\rho^5},
\end{equation}
here $\rho$ denotes the size of an instanton
and $\alpha_S = g^2(\rho)/(4\pi)$ the renormalized coupling constant.

At finite temperature the size of an instanton cannot be larger than
$1/(gT)$ because of Debye screening. Because large instantons cannot
contribute to the transition rate at large $T$ anymore and $g(T)$ goes
to zero due to asymptotic freedom, the transition rate due to
instantons will go down if the temperature is raised. More
specifically Pisarski and Yaffe \cite{Pisarski1980} computed the
transition rate in the absence of fermions and found
\begin{equation}
\frac{ \mathrm{d} \nt^\pm}
{\mathrm{d}^3 x \, \mathrm{d}t\, \mathrm{d}\rho} 
=
\exp(-2 \pi^2 \rho^2 T^2 - 18 A(\pi \rho T) )
\left. \frac{ \mathrm{d} \nt^\pm}
{\mathrm{d}^3 x \, \mathrm{d}t\, \mathrm{d}\rho}\right \vert_{T=0},
\end{equation}
where $A(\lambda)$ is a function
which behaves as $-\log(\lambda)/6$ for $\lambda \rightarrow
\infty$ and as $-\lambda^2 / 36$ for $\lambda \rightarrow 0$ 
\cite{Gross1980}.

Hence for high temperatures the transition rate due to instantons
becomes extremely small. This also happens in the electroweak theory
where the coupling constant is always tiny. In this case transitions
between different vacua lead to violation of baryon plus lepton
($B+L$) number \cite{thooft1976}.  However, it was realized that the
electroweak theory contains static solutions which have finite
energy and half integer winding number \cite{Klinkhamer1984}.  Such a
solution with minimal energy is called a sphaleron (ready to fall).
If the temperature is higher than the energy of the sphaleron it is
likely that one can go over the barrier instead of tunneling. This
increases the transition rate and hence the rate for baryon number
violation enormously
\cite{Kuzmin1985,Shaposhnikov1987,Arnold1987,Arnold1988}.

It was realized that these sphaleron configurations also exist in QCD
\cite{McLerran1990}. The typical energy of the sphalerons in QCD is
$\Lambda_{\mathrm{QCD}} \approx 200\;\mathrm{MeV}$. Hence also in QCD
at high temperatures one can go over the barrier which leads to an
enormous increase in the transition rate. According to
Refs.~\cite{moore1997}-\cite{bodeker1999} the rate for SU(2)
Yang-Mills theory is
\begin{equation}
\frac{ \mathrm{d} \nt^\pm}
{\mathrm{d}^3 x \, \mathrm{d}t}  \sim 25.4 \alpha^5_W T^4.
\end{equation}
The rate has not been directly measured for QCD, but if we assume
$N_c$ scaling, we estimate the rate for QCD to be
\begin{equation}\label{sphrate}
 \frac{ \mathrm{d} \nt^\pm}
{\mathrm{d}^3 x \, \mathrm{d}t}  \equiv \Gamma^{\pm} \sim 192.8 \alpha^5_S T^4.
\end{equation}

There is a subtlety in these considerations for massless quarks.  For
massless quarks, the overlap between a state with zero Chern-Simons
charge, and one with one unit of charge vanishes,
\begin{equation}
<n_w = 1 \mid n_w = 0> = 0,
\end{equation}
since in the path integral representation for such a transition, the
fermion determinant has a zero mode and vanishes.  However, there is a
non-zero contribution for
\begin{equation}
<n_w = 1 \mid \overline \psi \psi \mid n_w = 0>
\end{equation} 
since in the path integral representation for this quantity, the zero
from the determinant is canceled by a singularity from the quark
propagator which represents the operator $\overline \psi \psi$.  The
overall expression is proportional to zero mode wavefunctions.  This
operator corresponds to a chirality flip, and as it should satisfies
the anomaly relation between the change of helicity and the change of
Chern-Simons charge.

The total rate of transition is of course the sum of the rates of the
lowering and raising transitions,
\begin{equation}
  \frac{ \mathrm{d} \nt}
{\mathrm{d}^3 x \, \mathrm{d}t} 
= \sum_\pm
 \frac{ \mathrm{d} \nt^\pm}
{\mathrm{d}^3 x \, \mathrm{d}t}.
\end{equation}

\section{The Chiral Magnetic Effect}
\label{sec:mechanism}
We will show in this section that gauge field configurations with
nonzero $\qw$ can separate charge in the presence of a background
(electromagnetic) magnetic field $B$. This we call the Chiral Magnetic
Effect. We consider the most ideal situation in which all quark masses
vanish. This should be a good approximation in the deconfined, chiral
symmetry restored phase.  First we consider the case with a very large
magnetic field, which means for us that $e B$ is much larger than the
momentum squared of all particles, i.e.\ $e B \gg p^2$. Then we will
study the effect of a moderate field.

\subsection{Large magnetic field}
Initially we assume that we have  a very small number of left and
right-handed fermions.  This is because our initial state has little
fluctuation in the helicities of particles.  This is a reasonable
assumption for an initial state typical of a Color Glass Condensate,
where the coupling is weak and the fluctuations should be generated by
quantum mechanical tunneling, and therefore suppressed as
$\exp(-2\pi/\alpha_S)$ \cite{mv,iancu,Kharzeev:2000ef}.  At later time,
sphaleron--like transitions should be possible since the system
becomes an ensemble of classical configurations with energy typically
greater than the barrier \cite{krasnitz2002}.

\begin{figure}[htb]
\begin{center}
\includegraphics[width=8cm]{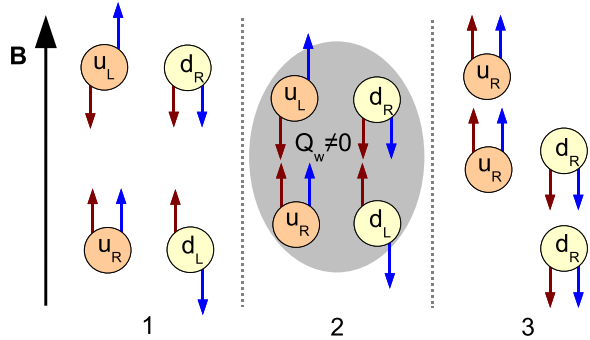}
\end{center}
\caption{Illustration of the Chiral Magnetic Effect in a very
  large homogeneous magnetic field.   The red arrows denote the direction
  of momentum, the blue arrows -- the spin of the quarks.  (1) Due to the
  very large magnetic field the up and down quarks are all in the
  lowest Landau level and can only move along the direction of the
  magnetic field.  Initially there are as many left-handed as
  right-handed quarks.  (2) The quarks interact with a gauge
  configuration with nonzero $\qw$.  Assuming $\qw = -1$, this will
  convert a left-handed up/down quark into a right-handed up/down
  quark by reversing the direction of momentum.  (3) The right-handed
  up quarks will move upwards, the right-handed down quarks will move
  downwards. A charge difference of $Q = 2 e$ will be created between
  two sides of a plane perpendicular to the magnetic field.  }
\label{fig:chargesep}
\end{figure}

 Since the magnetic field is so large, all particles will
be found in the lowest Landau level. Hence their spin is aligned along
the magnetic field and they can also only move along the magnetic
field.  Quarks with opposite charges have their spins aligned in
different directions.  If $B$ points in the $z$-direction positively
charged right-handed fermions and negatively charged left-handed
fermions will move in the positive $z$-direction (upwards). At the
same time positively charged left-handed fermions and negatively charged
right-handed fermions will move downwards.  We illustrated this
situation in Fig.~\ref{fig:chargesep}.

The fermions will then interact with a gauge field configuration, so
that some of them will change their helicity.  After the interaction
with the gauge field configuration has taken place we find that (since
we assumed the chiral limit)
\begin{equation}
(N^f_L - N^f_R)
= 2 \qw,
\end{equation}
so that if $\qw$ is nonzero we obtain a difference between the number
of left and right-handed fermions that is the same for each flavor.
The fermions can only change helicity by reversing their momenta,
since spin flip is energetically suppressed in a large magnetic field
(we assumed $e B \gg p^2$). So after the interaction has taken place
all positively charged right-handed fermions and negatively charged
left-handed fermions will still move upwards. At the same time
positively charged left-handed quarks and negatively charged
right-handed quarks will still move downwards.  However, now there is
a difference between the number of right-handed and left-handed
fermions. As a result an electromagnetic current is generated along
the direction of $B$. If we assume that initially all fermions are
within a finite volume, this current will induce a charge difference
$Q e$ between opposite sides a plane perpendicular to the magnetic
field. Here $e$ denotes the elementary charge and 
\begin{equation}
 Q = 2 \qw \sum_f \vert q_f \vert,
\label{eq:deltaq}
\end{equation}
where $q_f$ is the charge in units of $e$ of a quark with flavor $f$.
For $N_f=2$ and $N_f=3$ the relation above
becomes $ Q = 2 Q_w$ and $ Q = \tfrac{8}{3} Q_w$ respectively.
An alternative derivation of this result is presented in Appendix B.

Since Eq.~(\ref{eq:deltaq}) was obtained in the most ideal
case, i.e. chiral limit and an extremely large magnetic field,
Eq.~(\ref{eq:deltaq}) has to be an upper limit. Therefore 
there is a maximum amount of charge that can be separated
by a particular gauge field configuration in a homogeneous
background magnetic field; in other words
\begin{equation}
 \vert Q \vert \leq 2 \vert \qw \vert \sum_f \vert q_f \vert.
\label{eq:deltaqup}
\end{equation}

In our derivation we ignored the back-reaction due to the electric
field created by the separating charges. We believe that this 
back-reaction can only give rise to a small suppression of the effect. 
This is because the size of the electric field will be much smaller 
than $eB$ since there are very few particles involved. Moreover,
in the physical case we are interested in, a heavy ion collision, 
the color forces will surely dominate.

Since the separation is independent of color, the mechanism will
not create a net color charge difference. Therefore we can safely 
ignore a gluonic back-reaction.

\subsection{Moderate Magnetic Field}
Now that we considered the ideal situation with an extremely large
magnetic field, let us estimate the amount of charge separated in a
moderate homogeneous magnetic field. There are now three important
scales in the problem, $eB$, the temperature $T$, and the size $\rho$ of
the configuration with nonzero $\qw$.

In a moderate magnetic field not all spins will be aligned along
$B$. Let us introduce the function $\beta_f(\omega)$ which denotes the
degree of polarization along $B$ of quarks with flavor $f$ and energy
smaller than $\omega$.  Since a magnetic field does not distinguish
between left and right-handed particles, $\beta_f(\omega)$ will be the
same for left and right-handed particles in thermodynamic
equilibrium. Moreover, ignoring chemical potential, $\beta_f(\omega)$
will be equal for particles and antiparticles.

In a homogeneous magnetic field it is possible to calculate
$\beta_f(\omega)$. The energy dispersion relation of a massless
fermion is then given by
\begin{equation}
\omega_p^2 = p^2_3 + 2 \vert q_f e B \vert n,
\end{equation}
where $n=0,1,2,\ldots$ denotes the Landau level.  The $n=0$ level
contains only one spin direction, all other levels contain both spin
directions and are hence double degenerate.  

The number density of quarks in a magnetic field with energy smaller
than $\omega$ reads
\begin{equation}
 N_{\uparrow}  + N_{\downarrow} =
 \frac{ \vert q_f eB \vert}{2\pi^2} \sum_{n=0}^{\infty}
 \int_0^{\infty} \mathrm{d}p_3
 \alpha_n \theta(\omega^2 - \omega_p^2) n(\omega_p),
\end{equation}
where $\alpha_n = 2-\delta_{n,0}$ is the degeneracy factor and
$n(\omega_p) = [\exp(\omega_p/T) + 1]^{-1}$ is the Fermi-Dirac
distribution function.  The difference between the number density of
fermions with spin parallel and antiparallel to the magnetic field
yields
\begin{equation}
 \vert N_{\uparrow} - N_{\downarrow}\vert =
 \frac{ \vert q_f eB \vert}{2\pi^2}  
 \int_0^{\infty} \mathrm{d}p_3
 \theta(\omega^2 - p_3^2) n(p_3).
\end{equation}
Hence the degree of polarization of quarks/antiquarks with energy
smaller than $\omega$ is given by
\begin{equation}
 \beta_f(\omega) = \frac{\vert N_{\uparrow} -  N_{\downarrow} \vert}
{ N_{\uparrow}  + N_{\downarrow}}.
\label{eq:polar}
\end{equation}

Let us try to find an approximation for $\beta_f(\omega)$. For $2 \vert q_f
e B \vert > \omega^2$, $\beta_f(\omega) = 1$ since only the lowest Landau level
contributes.  For $2 \vert q_f e B\vert  = 0$, $\beta_f(\omega)=0$.  Therefore
\begin{equation}
 \beta_f(\omega) \approx \gamma \left( \frac{2 \vert q_f e B \vert}{\omega^2} \right)
\label{eq:betaf}
\end{equation}
with
\begin{equation}
 \gamma(x) = 
\left \{ \begin{array}{cl}
x & \mathrm{for}\; x \leq 1 \\
1 & \mathrm{for}\; x \geq 1
\end{array}
\right. .
\end{equation}
We have computed Eq.~(\ref{eq:polar}) numerically and found that
Eq.~(\ref{eq:betaf}) is a reasonable approximation.

At first hand, it might seem surprising that the temperature drops out
the estimate for the degree of polarization. The reason for this is
that we only considered the degree of polarization of a subset of all
modes, namely the ones with energy smaller than $\omega$. If one would
instead be interested in the polarization of the bulk of the
particles, one should take $\omega \approx T$, by which the
temperature dependence naturally is recovered.

Quarks with momenta much larger than the inverse size $1/\rho$ of the
configuration with nonzero winding number will not be affected by
these configurations and hence do not change helicity. Therefore the
quarks that change helicity will have momenta and hence energy smaller
than $\sim 1/\rho$.  Assuming that all quarks with energy smaller than
$1/\rho$ are equally likely to change helicity we find the following
estimate for the expectation value of the amount of charge separated
\begin{equation}
 Q \approx 2 \qw \sum_f \vert q_f \vert \gamma(2 \vert q_f \Phi \vert).
\label{eq:chargesepmodfield}
\end{equation}
where $\Phi = e B \rho^2$ is essentially the flux through a configuration
with nonzero $\qw$. For small fields ($2 \vert q_f e B \vert < \rho^2$) this reduces to
\begin{equation}
 Q \approx 4 \Phi \qw \sum_f q_f^2.
\end{equation}

\section{Chiral Magnetic Effect in Heavy Ion Collisions}
\label{sec:chiralmaghic}
Now that we know how much charge is separated by a single
configuration with winding number $\qw$, let us estimate the amount of
charge separated in a heavy ion collision. We will assume that hot
deconfined chirally restored matter is created, so our results of the
previous section will be applicable.

\begin{figure}[t]
\begin{center}
\includegraphics{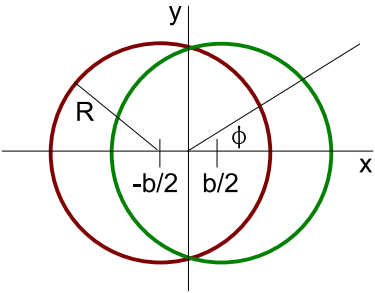}
\end{center}
\caption{Crossectional view of a non-central heavy ion collision along
  the beam-axis ($z$-axis). The two nuclei have radii $R$, travel in
  opposite directions and collide with impact parameter $b$.  The
  plane $y=0$ is called the reaction plane. The angle $\phi$ is an
  azimuthal angle with respect to the reaction plane. The region in
  which the two nuclei overlap contains the participants, the regions
  in which they do not contain the spectators.}
\label{fig:collision}
\end{figure}

As we have shown in the previous section, if $e B \sim 1/\rho^2$ there
is a good chance of getting charge separation. In a hot plasma, the
typical size of a configuration with nonzero $\qw$ is of the order of
the magnetic screening length, $\alpha_S T$.  The typical required
magnetic fields are therefore of order $\alpha_S^2 T^2$, which we
estimate to be of order $10^2 - 10^3 ~\mathrm{MeV}^2$, for the initial
conditions typical at RHIC.  In Appendix \ref{app:magfield}, we show
that such enormous magnetic fields can indeed be created in
off-central heavy ion collisions. In the plasma, the magnetic field is
to a good degree homogeneous and pointing in a direction perpendicular
to the reaction plane ($y = 0$ in Fig.~(\ref{fig:collision})) in the
region of overlap of the colliding nuclei for the impact parameters of
interest.

We will now define $\na^\pm$ and $\nb^\pm$ to be the total
positive/negative charge in units of $e$ above ($a$) and below ($b$) the reaction
plane respectively. 
 We have that $N^\pm = \na^\pm + \nb^\pm$, where
$N^\pm$ is the total positive/negative charge produced in a certain
event. We define $\Delta_{\pm}$ to be the difference between in charge
on each side of the reaction plane $\Delta_{\pm} = \na^\pm - \nb^\pm$.

One generates an electric current along the magnetic field each time a
transition from one vacuum to another is made.  Let us now imagine
such transition happening in the center of the hot matter. Locally a
charge difference will be created due to the current. But since the
quarks creating this current will still have many interactions before
they leave the hot matter, presumably transitions in the center due to
the final--state randomization of quark momenta will not give rise to
any observable charge separation.

Now let us look to a transition near the surface of the hot
matter. Suppose that the situation is such that the charge separation
mechanism causes more positively charged quarks to move outwards and
more negatively charged quarks the move inwards (the opposite
situation is of course also possible with an equal probability). The
positively charged quarks moving out of the surface will hadronize and
can essentially freely propagate to the detector.  On the other hand
the negatively charged quarks will move inwards and still encounter
many interactions. As a result the azimuthal angles of the positively
and negatively charged quarks will become uncorrelated. This is very
similar to the observed jet quenching in heavy ion collisions. 
Only for very large impact parameters, when the size of
the interacting region is about same as the scattering length, we
expect correlations between positively and negatively charged
particles to exist.

Each time a transition is made both $\Delta_+$ and $\Delta_-$ will
change. The expectation value of the change is either positive or negative with equal probability and
given by
\begin{equation}
  \pm \sum_f  \vert q_f \vert \ \gamma(2 \vert q_f \Phi \vert) \xi_\pm(x_\perp).
 \label{eq:deltachange}
\end{equation}
Here we used that the most probable transitions are those which
connect two neighboring vacua, so $\qw = \pm 1$.  The transitions
which simultaneously induce changes with $\qw > 1$ are suppressed,
therefore we have neglected them.  We compute those transitions which
are independent of one another, but taken over the time and volume of
the system can induce multiple changes in the Chern-Simons charge.
The functions $\xi_\pm(x_\perp)$ are screening suppression functions
in order to incorporate the effect that transitions near the surface
are much more likely to contribute to $\Delta_\pm$. We will take
\begin{equation}
 \xi_\pm(x_\perp) = \exp(-\vert y_\pm(x) - y\vert / \lambda),
 \label{eq:xipm}
\end{equation}
where $\lambda$ is the screening length. The functions $y_+(x)$ and
$y_-(x)$ denote the upper and lower $y$ coordinate of the overlap region. Hence
$\vert y_{\pm}(x) - y \vert$ is the distance from a point $y$ to the upper/lower
part of the surface. It follows from  Fig.~(\ref{fig:collision}) that
\begin{equation}
y_+(x) = -y_-(x) = \left \{ \begin{array}{lc} 
\sqrt{R^2 - (x-b/2)^2} & \hspace{0.6cm} -R +b/2 \leq x \leq 0
\\
\sqrt{R^2 - (x+b/2)^2} & \hspace{0.6cm}
0 \leq x \leq R -b/2
\end{array}
\right. ,
\end{equation}
here $R$ denotes the radius of the nuclei and $b$ is the impact
parameter.  Now one can easily infer from Eq.~(\ref{eq:deltachange})
that the screening works. It is equally likely to increase
$\Delta_\pm$ as to decrease it. However if for example $y$ is near
$y_+$ the expectation value for increasing $\Delta_\pm$ is much larger
then the expectation value for decreasing it.  Let us mention here
that Eq.~(\ref{eq:deltachange}) is strictly speaking only valid for a
constant homogeneous magnetic field. In the overlap region the
magnetic field is to a good degree homogeneous around zero space-time
rapditiy especially for large impact parameters. It has however a huge
time dependence. So in reality there could be effects which are due to
the rapid time variation of the field that we will ignore in this
paper.

Now we define $\nt^\pm$ to be the total number of raising/lowering
transitions. It of course holds that $\nt = \nt^+ +
\nt^-$. Furthermore we introduce $\dt$ which is the difference between
the number of raising and lowering transitions, i.e.~ $\dt = \nt^+ -
\nt^-$.  We can assume that all transitions happen independently from
each other. Then the dynamics for $\dt$ is exactly governed by a
one-dimensional random walk.  At each transition one has equal
probability to go up or down. Hence the expectation value of $\dt$
will vanish. But since there are $\nt$ transitions its variation is
equal to $\sqrt{\nt}$, so that
\begin{equation}
\langle \dt^2 \rangle = \int_{t_i}^{t_f} \mathrm{d} t\,
\int_V \mathrm{d}^3 x
\int \mathrm{d} \rho
\frac{ \mathrm{d} \nt}
{ \mathrm{d}^3 x\, \mathrm{d}t\, \mathrm{d}\rho},
\end{equation}
where $V$ denotes the volume in which the transitions take place.

Now we can compute the variation of $\Delta^\pm$. Since we assume that
all transitions are independent from each other the total variation is
the sum of the variation of all contributions. Hence it follows that
\begin{multline}
 \langle \Delta_\pm^2 
 \rangle 
= 
\tfrac{1}{2} \int_{t_i}^{t_f} \mathrm{d} t\,
\int_{V} \mathrm{d}^3 x
\int \mathrm{d} \rho\,
\frac{ \mathrm{d} \nt}
{ \mathrm{d}^3 x\, \mathrm{d}t\, \mathrm{d}\rho}
\left[\xi_-(x_\perp)^2 + \xi_+(x_\perp)^2 \right]
\times 
\\
\left[ \sum_f 
\vert q_f \vert \gamma(2 \vert q_f e B \vert \rho^2) \right]^2.
\label{eq:deltapm}
\end{multline}
Likewise we can compute $\langle \Delta_+ \Delta_-\rangle$. We find
\begin{multline}
 \langle \Delta_+ \Delta_- 
 \rangle 
= 
- \int_{t_i}^{t_f} \mathrm{d} t\,
\int_{V} \mathrm{d}^3 x
\int \mathrm{d} \rho\,
\frac{ \mathrm{d} \nt}
{ \mathrm{d}^3 x\, \mathrm{d}t\, \mathrm{d}\rho}
\xi_-(x_\perp) \xi_+(x_\perp)
\times 
\\
\left[ \sum_f 
\vert q_f \vert \ \gamma(2 \vert q_f e B \vert \rho^2) \right]^2.
\label{eq:deltapdeltam}
\end{multline}
Let us remark here that if screening is absent we have $\langle
\Delta_\pm^2 \rangle = -\langle \Delta_+ \Delta_- \rangle$ since then
$\xi_\pm = 1$. In the presence of screening however, $\xi_+ \xi_- < 
\frac{1}{2} (\xi_+^2 + \xi_-^2)$ so that $\vert 
\langle \Delta_+ \Delta_-\rangle \vert < \langle \Delta_\pm^2 \rangle$.

Using the fact that $\rho \sim (\Gamma^{\pm}/\alpha_S)^{-1/4} \sim
1/(\alpha_S T)$, and the expression for the sphaleron transition rate
Eq.~(\ref{sphrate}) we can rewrite the last two formula's for small magnetic
fields ($2 \vert q_f e B \vert < 1/\rho^2$) as
\begin{equation}
 \frac{\mathrm{d} \langle \Delta_\pm^2 
 \rangle }{\mathrm{d} \eta}
= 2 \kappa\ \alpha_S \left[ \sum_f q_f^2 \right]^2
\int_{V_{\perp}} \mathrm{d}^2 x_\perp \left[ \xi_+^2(x_\perp) + \xi_-^2(x_\perp) \right]
\int_{\tau_i}^{\tau_f} \mathrm{d} \tau \tau \, 
[e B(\tau, \eta, x_\perp) ]^2,
\end{equation}
\begin{equation}
 \frac{\mathrm{d} \langle \Delta_+ \Delta_- 
 \rangle }{\mathrm{d} \eta}
= - 4 \kappa\ \alpha_S \left[ \sum_f q_f^2 \right]^2
\int_{V_{\perp}} \mathrm{d}^2 x_\perp \xi_+(x_\perp) \xi_-(x_\perp)
\int_{\tau_i}^{\tau_f} \mathrm{d} \tau \tau \, 
[e B(\tau, \eta, x_\perp) ]^2,
\end{equation}
where the proper time $\tau = (t^2 - z^2)^{1/2}$ and the space-time
rapidity $\eta = \frac{1}{2} \log[ (t + z)/(t-z)]$. The volume integral
is over the overlap region $V_\perp$ in the transverse plane. The
time integral is from the initial time $\tau_i$ to a final time $\tau_f$.
We have assumed
here that the magnetic field does not change the sphaleron transition
rate dramatically.  We have also inserted a constant $\kappa$ which
should be of order one, but with large uncertainties, since in our
estimate of the scale size there is an uncertainty of order one which
is taken to the fourth power.  We assumed that the scale size factor
in the sphaleron roughly canceled the factor of $385 \alpha_S^4$ in
the numerator, since the rate per unit volume should be of the order
of the inverse size of the typical field configuration to the fourth
power. 

At this point, we could take our exact expression for the magnetic
field given in Appendix A and compute the rate. This we will do in a
later paper for a variety of initial conditions. For now however 
we will use the approximate formulae for the magnetic field at 
the origin due to
spectators and participants. The approximations are given in
Eqs.~(\ref{eq:magfieldapproxspect}) and (\ref{eq:magfieldapprox}).
and valid for $R / \sinh(\rap_0) \lesssim \tau \lesssim R$. Adding
the contributions of the spectators and participants we find
\begin{equation}
 e B \approx Z \alpha_{EM} \left[c \exp(-\rap_0/2) \frac{1}{R^{1/2} \tau^{3/2}} 
f(b/R)
+ 
4 \exp(-2 \rap_0) \frac{b}{\tau^3} \right]
.
\label{eq:magfieldtotalapprox}
\end{equation}
Here $Z$ denotes the charge of the nucleus, $R$ its radius and
$\rap_0$ the beam rapidity. The function $f(b/R)$ given in
Eq.~(\ref{eq:impactpardep}) and displayed in Fig.~(\ref{fig:f})
describes the impact parameter dependence of the participant
contribution to the magnetic field to first approximation. The
constant $c = 0.59907\ldots$ and $\alpha_{EM} \approx 1/137$ denotes
the fine structure constant. Especially for large impact parameters
the magnetic field at the origin is very homogeneous in the transverse
plane within the overlap region, so that
Eq.~(\ref{eq:magfieldtotalapprox}) is a good approximation to the
field at the surface of the interacting matter. For smaller impact
parameters the magnetic field at the surface is somewhat smaller, so
that we might overestimate the magnetic field at the surface using
Eq.~(\ref{eq:magfieldtotalapprox}) for small impact parameters.

Let us assume that the screening length $\lambda$ is 
constant in this short time.  Then the integral over the
transverse plane can be done since we used that
the magnetic field is homogeneous to a certain degree. 
We define
\begin{equation}
\begin{split}
g(b/R, \lambda/R) \equiv &  \frac{1}{2 R^2} \int_{V_\perp} \mathrm{d}^2 x_\perp 
\left[ \xi_+^2(x_\perp) + \xi_-^2(x_\perp) \right],
\\
h(b/R, \lambda/R) \equiv & \frac{1}{R^2}
\int_{V_\perp} \mathrm{d}^2 x_\perp 
\xi_+(x_\perp)\xi_-(x_\perp).
\end{split}
\label{eq:gh}
\end{equation}
In the limit of $\lambda \rightarrow \infty$ these
functions become equal to
\begin{equation}
\frac{1}{R^2} \int_{V_{\perp}} \mathrm{d}^2 x_\perp =
2 \arccos \left( \frac{b}{2R} \right) - \frac{b}{R} \sqrt{1 - \frac{b^2}{4 R^2}}.
\label{eq:vperp}
\end{equation}

We can now perform the integration over proper time. In the
integration over proper time, the magnetic field is sharply
falling. We can therefore to good approximation take $\tau_f = \infty$
and find
\begin{multline}
 \frac{\mathrm{d} \langle \Delta_\pm^2 
 \rangle }{\mathrm{d} \eta}
= 4 Z^2 \alpha_{EM}^2 \kappa\ \alpha_S \left[ \sum_f q_f^2 \right]^2
g(b/R, \lambda/R)
\left[ c^2 \exp(-\rap_0) f^2(b/R) \frac{R}{\tau_i}
\right.
\\
\left.
+ \frac{16}{5} c \exp(-5/2 \rap_0)f(b/R) \frac{b R^{3/2}}{\tau_i^{5/2}}
+
4 \exp(-4\rap_0) \frac{b^2 R^2}{\tau_i^4}
\right].
\end{multline}

We now must motivate a choice of initial time in the above equation.
If the time at which the topological charge changing processes is less
than the Lorentz contracted size of the system, then the magnetic
field is at its maximum value, and our approximation of treating the
spectator nucleons as being on an infinitesimally thin sheet breaks
down.  At the time when this approximation breaks down, the field
should have achieved its maximum value.  In this case, we should be
able to approximate the initial time as
\begin{equation} 
	\tau_i = \zeta R~ e^{-Y_0}
\end{equation}
where $\zeta \sim 2$.
On the other hand, if the initial time for the topological charge changing processes is larger
than this, we should be able to approximate
\begin{equation}
          \tau_i \sim 1/Q_{sat}
\end{equation}
where $Q_{sat}$ is the saturation momentum.  The saturation momentum
sets the time scale for the evolution of classical fields produced in
the collisions.  The choice between these two times scales is the
maximum of the two possible values.

For gold nuclei at $100~\mathrm{GeV}$ per nucleon, both these choices
are roughly the same and of the order of $0.1 -
0.2~\mathrm{fm/c}$. For copper at the same energy, the choice would be
the inverse saturation momentum since a copper nucleus is much smaller
than gold.  For gold at lower energy, we would best choose the Lorentz
contracted size scale, since the finite size of the nucleus becomes
important.

If we use the Lorentz contracted size scale, appropriate for gold, then
\begin{multline}
\frac{\mathrm{d} \langle \Delta_\pm^2 
 \rangle }{\mathrm{d} \eta}
= 4 Z^2 \alpha_{EM}^2 \kappa\ \alpha_S \left[ \sum_f q_f^2 \right]^2
g(b/R, \lambda/R)
\left[ c^2 f^2(b/R) \frac{1}{\zeta} 
\right.
\\
\left.
+ \frac{16}{5} c f(b/R) \frac{b}{R} \frac{1}{\zeta^{5/2}}
+
4 \frac{b^2}{R^2} \frac{1}{\zeta^{4}}
\right].
\label{eq:avgdeltapsqfinal}
\end{multline}
Similarly we find
\begin{multline}
\frac{\mathrm{d} \langle \Delta_+ \Delta_- 
 \rangle }{\mathrm{d} \eta}
= - 4 Z^2 \alpha_{EM}^2 \kappa\ \alpha_S \left[ \sum_f q_f^2 \right]^2
h(b/R, \lambda/R)
\left[ c^2 f^2(b/R) \frac{1}{\zeta} 
\right.
\\
\left.
+ \frac{16}{5} c f(b/R) \frac{b}{R} \frac{1}{\zeta^{5/2}}
+
4 \frac{b^2}{R^2} \frac{1}{\zeta^{4}}
\right].
\label{eq:avgdeltapmfinal}
\end{multline}

We expect that taking $\zeta = 2$ is reasonable. 
For two flavors we get $(\sum_f q_f^2)^2 = 25/81$.  
Then we find for collisions at large impact parameters
when screening is unimportant
\begin{equation}
\frac{\mathrm{d} \langle \Delta_\pm^2 
 \rangle }{\mathrm{d} \eta} \sim 
\frac{100}{81} Z^2 \alpha_{EM}^2 \alpha_S \kappa
\left[2 \arccos\left(\frac{b}{2 R} \right)
- 
\frac{b}{R} \sqrt{1 - b^2 / 4 R^2} 
\right] \frac{b^2}{4 R^2}.
\end{equation} 
Now let us take $b/R = 1.6$ which roughly corresponds to $50-60 \%$
centrality. For gold-gold collisions ($Z=79$ and $E/A =
100\;\mathrm{GeV}$ ) we then find taking $\alpha_s = 1$ and $\kappa =
1$
\begin{equation}
 \frac{\mathrm{d} \langle \Delta_\pm^2 
  \rangle }{\mathrm{d} \eta} \sim  0.1. 
 \label{eq:preddeltasq}
\end{equation} 
Let us stress again that this is a very rough estimate. For example
taking the initial time twice as large will decrease our estimate
by a factor 16. We could easily increase our estimate by an order of magnitude
by taking $\kappa = 10$.

In the next section we will connect $\langle \Delta^2_\pm \rangle$ and
$\langle \Delta_+ \Delta_- \rangle$ to some observables. Then we will
discuss in detail the implications of Eqs.~(\ref{eq:avgdeltapsqfinal}) and
(\ref{eq:avgdeltapmfinal}).
 
\section{Observables} 
Due to the finite multiplicity of produced particles, one cannot
observe charge separation in an individual event. The statistical
fluctuations $\sim \sqrt{N}$ will be much larger than the expected
charge asymmetry. By taking an average over many events $\Delta_\pm$ will of
course vanish, since configurations with positive and negative $\qw$
are very likely to be produced with equal probability.  However can
define correlators which are sensitive to the variation $\langle 
\Delta_\pm^2 \rangle$. In this section we are going to connect
these correlators to $\langle \Delta_\pm^2 \rangle$.

A set of very useful correlators for the study of charge separation
was proposed by Voloshin~\cite{Voloshin:2004vk}. For each event one defines
\begin{equation}
f(\phi_a, \phi_b)
= \frac{1}{N_a N_b} \sum_{i=0}^{N_a} \sum_{j=0}^{N_b} \cos(\phi_{ai} + \phi_{bj}),
\end{equation}
where $a, b = \pm$ denotes the charge,  $N_\pm$ is the total number of
positively or negatively charged particles, and $\phi_{a i}$ denotes the
azimuthal angle of an individual charged particle with respect to the
reaction plane.

Then in order to remove the multiplicity fluctuations one averages the
correlators over $N_{\mathrm{e}}$ similar events. This averaging is
called event mixing. In this way one can define the averaged
correlators $a_{--}$, $a_{++}$ and $a_{+-}$ \cite{Voloshin:2004vk}, where
\begin{equation}
 a_{ab} = -\frac{1}{N_{\mathrm{e}}} \sum_{n=1}^{N_\mathrm{e}}
f(\phi_a, \phi_b).
\end{equation}

We feel that it is also useful to define correlators which
are not divided by square of the total multiplicity of charged
particles. Here we introduce $b_{--}$, $b_{++}$ and $b_{+-}$, where
\begin{equation}
 b_{ab} = -\frac{1}{N_{\mathrm{e}}} \sum_{n=1}^{N_\mathrm{e}}
g(\phi_a, \phi_b),
\end{equation}
with
\begin{equation}
g(\phi_a, \phi_b)
=  \sum_{i=0}^{N_a} \sum_{j=0}^{N_b} \cos(\phi_{ai} + \phi_{bj}).
\end{equation}
The correlators can also be written as
\begin{equation}
 g(\phi_a, \phi_b) = X_a X_b - Y_a Y_b, 
\end{equation}
where
\begin{equation}
 X_\pm \equiv 
\sum_{i=1}^{N_\pm}  \cos \left(
\phi_{i}^{\pm} \right ), 
\;\;\;\;\;\;\;\;\;
 Y_\pm \equiv
\sum_{i=1}^{N_\pm}  \sin \left(
\phi_{i}^{\pm} \right ).
\end{equation}
An equivalent representation of the last
two equations reads
\begin{equation}
 X_\pm =
 \int_0^{2 \pi} \mathrm{d} \phi \, \cos \left(
\phi \right ) \frac{ \mathrm{d} N_\pm} {\mathrm{d} \phi},
\;\;\;\;\;\;\;\;
 Y_\pm 
=
\int_0^{2\pi} \mathrm{d} \phi \, 
\sin \left( \phi \right )
\frac{ \mathrm{d} N_\pm} {\mathrm{d} \phi}
,
\end{equation}
where $\mathrm{d} N_\pm$ denotes the number of charged particles in
the azimuthal angle $\mathrm{d} \phi$.  Now one can see that $X_\pm$
essentially measures directed flow.  By taking a symmetric interval
in rapidity this quantity vanishes.  On the other hand $Y_\pm$
measures the difference between charges on opposite sides of the
reaction plane.  Let us emphasize that if a measurement shows that
$a_{++}/b_{++}$ is positive there really is on average more charge on
one particular side of the reaction plane.  Possible contamination by
directed flow will always give a negative contribution to
$a_{++}/b_{++}$.  The correlator $a_{+-}/b_{+-}$ measures correlations
between positively and negatively charged particles and the reaction
plane.  If such correlation exists and positively charged particles
are emitted oppositely from negatively charged particles
$a_{+-}/b_{+-}$ is necessarily negative.

In order to relate the correlators to $\Delta_\pm$ we have to make an
assumption of the distribution of the charges in the azimuthal angle.
If all charged particles affected by the Chiral Magnetic Effect
would be really emitted perpendicular to the
reaction plane in which case $\phi = \pi/2$ or $\phi = 3\pi/2$ we
would have
\begin{equation}
 Y_\pm = \Delta_\pm.
\end{equation}
As a result
\begin{equation}
  a_{++}  = 
 a_{--}  = 
  \frac{1}{N_+^2} \langle \Delta^2_\pm \rangle,
\;\;\;\;\;\;
a_{+-} = \frac{1}{N_+ N_-} \langle \Delta_+ \Delta_- \rangle,
\end{equation}
and
\begin{equation}
  b_{++}  = 
 b_{--}  = 
  \langle \Delta^2_\pm \rangle,
\;\;\;\;\;\;
b_{+-} = \langle \Delta_+ \Delta_- \rangle.
\end{equation}
where $N_\pm$ denotes the total number of charged particles in the
corresponding $\eta$ interval.

However, it might be more realistic that the particles will spread
along the direction of the average magnetic field.  If for example
\begin{equation}
 \frac{ \mathrm{d} N_\pm} {\mathrm{d} \phi} 
  = 
  \frac{1}{2\pi} N_\pm  + \frac{1}{4} \Delta_\pm \sin (\phi),
\end{equation}
we would find
\begin{equation}
 Y_\pm = \frac{\pi}{4} \Delta_\pm.
\end{equation}
In that case
\begin{equation}
  a_{++}  = 
 a_{--}  = 
  \frac{1}{N_+^2}   \frac{\pi^2}{16} \langle \Delta^2_\pm \rangle,
\;\;\;\;\;\;
a_{+-} = \frac{1}{N_+ N_-}   \frac{\pi^2}{16} \langle \Delta_+ \Delta_- \rangle.
\end{equation}
Similarly we find
\begin{equation}
 b_{++}  = 
 b_{--}  = 
   \frac{\pi^2}{16}
 \langle \Delta^2_\pm \rangle,
\;\;\;\;\;\;
b_{+-} = 
  \frac{\pi^2}{16}
\langle \Delta_+ \Delta_- \rangle.
\end{equation}

\section{Discussion}
One should be aware that the observables defined in the previous
section are $\mathcal{P}$ and $\mathcal{T}$ even.  For vanishing net charge density per unit rapidity at $\eta = 0$, the invariance under
$\mathcal{C}$ predicts that $a_{++}= a_{--}$ and $b_{++} = b_{--}$.
Hence not only the Chiral Magnetic Effect contributes to the observables, but
also some $\mathcal{P}$ even processes could do in principle. In order
to find out whether the signal seen in Ref.~\cite{Selyuzhenkov:2005xa}
is really due to the Chiral Magnetic Effect one has to perform a
detailed analysis and see whether calculations based on the Chiral
Magnetic Effect are consistent with the data, and rule out more mundane
explanations.  Based on our main
results Eqs.~(\ref{eq:avgdeltapsqfinal}) and
(\ref{eq:avgdeltapmfinal}) we therefore will make some predictions.
Let us make a cautionary statement here. We have made many
approximations and there are  many uncertainties. We expect that at
best we might describe the data only semi-quantitatively.

\subsection{Absolute Value of Asymmetry}
In Eq.~(\ref{eq:preddeltasq}) we estimated with 
large uncertainty $\langle \Delta_\pm^2 \rangle = 0.1$
at $b/R = 1.6$ which corresponds roughly to 50-60 \% centrality. 
Using that that  there are about
24 charged particles produced within $\vert \eta \vert < 0.5$ 
(see Table~\ref{table:impactcentrality}) we find $a_{++}=a_{--} 
\sim 10^{-4}$. Now this estimate is very uncertain, we can easily
increase or decrease it by an order of magnitude. However it
seems comparing this number to the results presented in
Ref.~\cite{Selyuzhenkov:2005xa} that our prediction has
about the same order of magnitude as seen in experiment.
The main lesson from this is that although
we are very uncertain about the value of the asymmetry,
the Chiral Magnetic Effect does not give rise to an asymmetry
which is many orders of magnitude away from the data obtained by experiment.

\subsection{$Z$ Dependence}
The $Z$ (charge/atomic number) is probably our most robust prediction.
Clearly the magnetic field is proportional to $Z$. Since the
polarization of the quarks is proportional to $B$, the correlators
$b_{ab}$ are proportional to $Z^2$. If polarization would become
maximal however, which we believe is unlikely, there will be no
$Z$-dependence.  In principle one could test $Z$-dependence by taking
nuclei with the same $A$ and different $Z$. This however could
be difficult since it will require very accurate data.

\subsection{$A$ Dependence}

The A dependence of our result depends upon which time scale is most
important, that associated with the Lorentz contracted size of the
nuclei, or that of the inverse saturation momenta.  The specific $A$
dependence depends upon what range of impact parameter, beam energy
and nuclear size we consider.

\subsection{Beam Energy Dependence}
We are very uncertain about the beam energy dependence, since our
result will depend strongly on what we take as our initial time. If
the initial time scales with the longitudinal size of the nucleus, $R
\exp(-\rap_0)$, we expect no beam-energy dependence at all for
$b_{ab}$.  However if it turns out to be better to use the inverse
saturation momentum for initial time, we can get a large
dependence. In such case we always expect (as long as the pancake
approximation stays good and a quark gluon plasma is formed) that
$a_{ab}$ and $b_{ab}$ are smaller at larger beam energies, we never
expect them to become larger.

\subsection{Centrality Dependence}
\begin{figure}[htb]
\begin{center}
\includegraphics{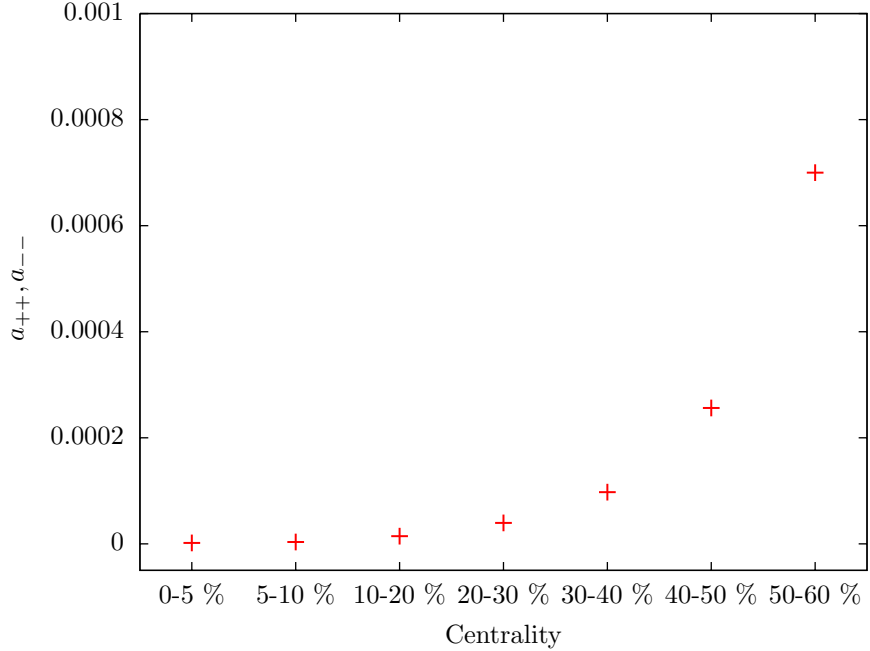}
\caption{Possible result of the Chiral Magnetic Effect for $a_{++}$ and $a_{--}$ as
a function of centrality, for Gold-Gold collisions at 130 GeV 
per nucleon pair.
\label{fig:appamm}
}
\end{center}
\end{figure}

The centrality dependence of the correlators $a_{ab}$ and $b_{ab}$ has a
strong dependence on initial time and could also have beam energy
dependence.  Let us assume that $\tau_i = \zeta R \exp(-\rap_0)$, we
can then use Eq.~(\ref{eq:avgdeltapsqfinal}) to determine the
centrality dependence of $a_{++}$, $a_{--}$ and $b_{--}$, $b_{++}$. For now
we will take $\zeta=2$ and $\lambda = 1\;\mathrm{fm}$ and present a plot
of $a_{ab}$ of our model for Gold-Gold. It could be very well argued
that one should take other values, but these seem to us not so
unreasonable. In order to compare our results with the data we will
plot $a_{++}$, $a_{--}$ as a function of centrality and normalize the
value in the 50-60~\% centrality bin to $7 \times 10^{-4}$
which is the value found by experiment~\cite{Selyuzhenkov:2005xa}. To convert
the impact parameter to centrality we used a table provided
in Ref.~\cite{nardi}. 
Using their results we also computed
the number of charged particles produced in each centrality
bin (see also Ref.~\cite{Adler2005} for experimental data).  
These results are summarized in Table~\ref{table:impactcentrality}.
Our result for $a_{++}$ and $a_{--}$ is displayed in Fig.~\ref{fig:appamm}.
Qualitatively the result agrees with the data presented in 
Ref.~\cite{Selyuzhenkov:2005xa}.

\begin{table}
\begin{center}
\begin{tabular}{c|c c c}
centrality & $b/R$ & $N_{\mathrm{part}}$ &$(N_+ + N_-)/2$ \\
\hline 
 0 - 5 \% &    0.35  & 344 &  283  \\
 5 - 10 \% &   0.47  & 317 &  256  \\
 10 - 20 \% &  0.82  & 222 &  169  \\
 20 - 30 \% &  1.07  & 153 &  111  \\
 30 - 40 \% &  1.26  & 102 &  72  \\
 40 - 50 \% &  1.43  & 64 &   43 \\
 50 - 60 \% &  1.58  & 38 &  24  \\
 60 - 70 \% &  1.72  & 20 &   12 \\
 70 - 80 \% &  1.84  & 9.5 &  5.5  \\
\end{tabular}
\end{center}
\caption{Relation between centrality, impact parameter, number 
of participants $N_{\mathrm{part}}$, and
the average number of produced charged particles $(N_+ + N_-)/2$ within $ \vert \eta \vert < 0.5$
for Gold-Gold collisions at 130 GeV per nucleon pair.
\label{table:impactcentrality}
}
\end{table}

\subsection{ Suppression of $a_{+-}$ Relative to $a_{++}$ and $a_{--}$}
\begin{figure}[t]
\begin{center}
\includegraphics{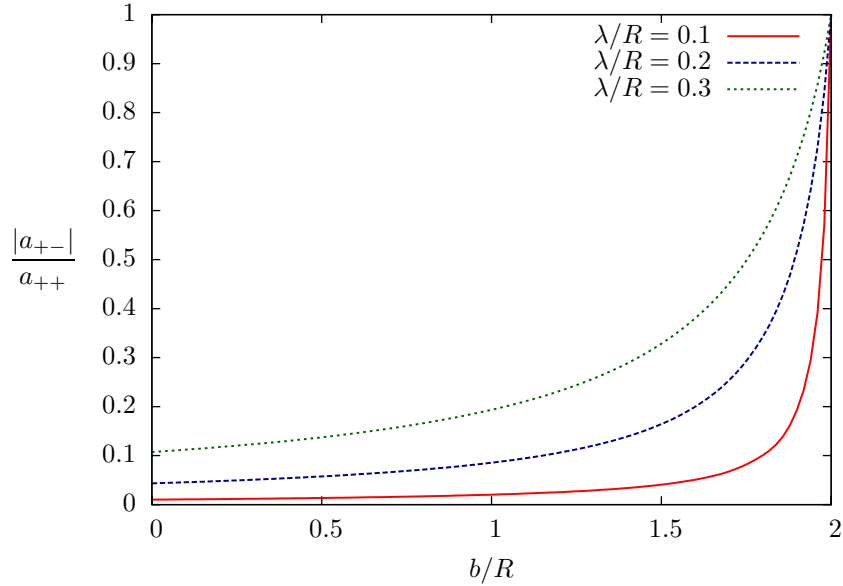}
\caption{Possible result of the correlator $\vert a_{+-} \vert / a_{++}$
as a function of $b/R$ for different values of the screening length
$\lambda/R$.
\label{fig:apmoverapp}}
\end{center}
\end{figure}

In Sec.~\ref{sec:chiralmaghic} we found that
$ \vert \langle \Delta_{+} \Delta_- \rangle \vert \leq
\langle \Delta_{\pm}^2 \rangle$ because of the screening
effect. As a result $\vert a_{+-} \vert \leq a_{++}$.
So it is very natural that the absolute value
of $a_{+-}$ is smaller than that of $a_{++}$.
It easily follows from our results that
\begin{equation}
 \frac{ \vert a_{+-} \vert }{a_{++}} 
 = \frac{h(b/R, \lambda/R)}{g(b/R, \lambda/R)},
\end{equation}
where the functions $g(b/R, \lambda/R)$ and $h(b/R, \lambda/R)$ are
given in Eq.~(\ref{eq:gh}).  We displayed $\vert a_{+-} \vert /
a_{++}$ as a function of $b/R$ in Fig.~(\ref{fig:apmoverapp}) for
different values of $\lambda/R$.  To compute the functions $g(b/R,
\lambda/R)$ and $h(b/R, \lambda/R)$ we used the screening suppression
factor $\xi_\pm(x_\perp)$ given in Eq.~(\ref{eq:xipm}).  It could very
well be that this simple suppression is to simplistic, and has to be
modified somehow. Especially near the surface, and at large impact
parameter, the function could be different. Nevertheless, one can
infer from Fig.~(\ref{fig:apmoverapp}) that for reasonable values of
$\lambda/R$ one finds an order of magnitude suppression of $a_{+-}$
compared to $a_{++}$ for small impact parameters which is
consistent with data presented in Ref.~\cite{Selyuzhenkov:2005xa}.
For larger impact parameters the suppression is smaller, since the
system size is in that case smaller.

Since the screening length $\lambda$ is independent of the size of the
system, we expect $\vert a_{+-} \vert /a_{++}$ and $\vert b_{+-} \vert
/b_{++}$ to be larger for smaller systems. This implies that $\vert
a_{+-}\vert /a_{++}$ and $\vert b_{+-}\vert /b_{++}$ should be larger
in Copper-Copper collisions than in Gold-Gold at the same centrality
and energy.

\subsection{Transverse Momentum Dependence}
We expect that the particles which make the helicity flip have
momentum in the order of $\alpha T$, hence we expect that the main
contribution to our $b_{ab}$ comes from particles which have low
transverse momentum.  However we cannot clearly determine what is low.
To make a safe prediction we expect the main contribution to $b_{ab}$
come from particles which have transverse momentum smaller than
$1\;\mathrm{GeV}$.

\subsection{Particle Species Dependence}
We think the main contribution to $b_{ab}$ should come from pions,
since the Chiral Magnetic Effect also creates an isospin difference,
but no baryon number difference if $N_f=2$ (see also Appendix
B). However at this point we cannot make definite statements about the
ratios protons to pions or kaons to pions that contributes to
$b_{ab}$.

\subsection{Order Parameter for Chiral Symmetry/Confinement?}
Since it is absolutely necessary for the Chiral Magnetic Effect to
operate that chiral symmetry is restored and matter is in the
deconfined phase, measurement of $a_{ab}/b_{ab}$ if due to the chiral
magnetic effect could also serve as an order parameter for chiral
symmetry and confinement. If the temperature (beam energy) is too low
such that there is chirally broken confined matter created we think it
is likely that $a_{ab}$ and $b_{ab}$ should vanish or at least become
much smaller.

\section{Summary}
The Chiral Magnetic Effect allows one to in principle probe the
effects of topological charge changing processes in QCD.  These
phenomena are of great interest in field theory, and of great
potential application in cosmology.  In QCD, such phenomena are
perhaps at the generation of nucleon mass, and also are associated
with dynamical breaking of $\mathcal{P}$ and $\mathcal{CP}$ symmetries.
 
A confirmed observation of such an effect would be of great import.  A
confirmed observation must also be subject to alternative hypothesis.
Our explanation has the advantage that many of its features may be
varied by choice of beam energy, and the Z and A of the colliding
nuclei.  The computations we present are at present semi-quantitative
estimates, that can be refined by more careful treatment of the
magnetic field, and a better computation of the underlying topological
charge changing processes.

\section*{Acknowledgments}
We would like to thank Vasily Dzordzhadze, Jianwei Qiu, Ilya
Selyuzhenkov, Yannis Semertzidis and Sergei Voloshin for
discussions. At various stages of this work we have benefited from
communications with Robert Pisarski, Mikhail Shaposhnikov, Edward
Shuryak and Ariel Zhitnitsky.  This manuscript has been authored under
Contract No.~\#DE-AC02-98CH10886 with the U.S.\ Department of Energy.

\appendix

\section{Magnetic Field in Heavy Ion Collisions}
\label{app:magfield}
Let us estimate the size of the magnetic field achieved in a heavy ion
collision. We consider two similar nuclei with charge $Z$ and radius
$R$.  These nuclei are traveling with rapidity $\rap_0$ in the
positive and negative $z$-direction. At $t=0$ they collide with impact
parameter $b$. We take the center of the nuclei at $t=0$ at $x = \pm
b/2$ so that the impact parameter direction lies along the $x$-axis. 
We depicted this situation in Fig.~\ref{fig:collision}.

The nuclei in typical heavy-ion collision experiments are nearly
traveling with the speed of light. For example gold-gold collisions
at RHIC are performed at center of mass energies of
$200\;\mathrm{GeV}$ per nucleon pair which corresponds to $\rap_0
\approx 5.4$. In that case the Lorentz contraction factor $\gamma$ is
equal to $100$.  Hence the nuclei are Lorentz contracted in the
$z$-direction to 1 percent of their original size. Therefore
to good approximation we can consider the two including nuclei as
pancake shaped. As a result we can assume that number densities of the
two nuclei are given by
\begin{equation}
 \rho_\pm(\vec x'_\perp) = \frac{2}{\tfrac{4}{3} \pi R^3}
 \sqrt{R^2 - ( \vec x'_\perp \pm \vec b / 2)^2}
\, \theta_\pm(\vec x'_\perp)
\end{equation}
here $\vec x'_\perp$ denotes a vector transverse to the 
beam axis $\vec{e}_z$ and
\begin{equation}
\theta_\pm(\vec x'_\perp) = \theta \left [R^2 - 
(\vec x'_\perp \pm \vec b / 2)^2 \right],
\end{equation}
The functions $\theta_\pm(\vec x'_\perp)$ are the projections of the nuclei
on the plane transverse to the beam axis and will also be used to separate
the spectators from the participants. The number densities are normalized such that
\begin{equation}
 \int \mathrm{d}^2 \vec x'_\perp \rho_\pm(\vec x'_\perp)
 = 1.
\end{equation}

First let us quote the magnetic field at position $\vec x = (\vec
x_\perp, z)$ caused by a particle with charge $Z$ moving in the
positive $z$-direction with rapidity $\rap$. At $t=0$ the particle can
be found at position $\vec x'_\perp$.  This magnetic field can be
obtained by either boosting the electric field of a charge $Z$ or
using the Li\'enard-Wiechert potentials. One finds ($e=\vert e \vert$)
\begin{equation}
 e \vec B(\vec x) = Z \alpha_{EM} \sinh (\rap) 
\frac{ (\vec x'_\perp - \vec x_\perp) \times \vec{e}_z}
{
\left[(\vec x'_\perp-\vec x_\perp)^2 + (t \sinh \rap - z
 \cosh \rap )^2 \right]^{3/2}
},
\end{equation}

Now we can estimate the strength of the magnetic field during a heavy
ion collision. We are only interested in $t>0$, i.e.\ just after the
collision. Then we can split the magnetic field in the following way
\begin{equation}
 \vec B = \vec B^+_s + \vec B^-_s + \vec B^+_p + \vec B^-_p,
\end{equation}
where $\vec B^\pm_s$ and $\vec B^\pm_p$ are the contributions of the
spectators and the participants moving in the positive or negative
$z$-direction respectively. We assume that the spectators do not
scatter at all and stay traveling with the beam rapidity $\rap_0$. It
follows that ($Y_0 > 0$)
\begin{multline}
 e \vec B^\pm_s(\tau, \eta, \vec x_\perp) = 
\pm Z \alpha_{EM} \sinh(\rap_0 \mp \eta) \int \mathrm{d}^2 \vec x'_\perp
\rho_\pm(\vec x'_\perp) [1 - \theta_\mp(\vec x'_\perp) ]
\\
\times
\frac{ (\vec x'_\perp - \vec x_\perp) \times \vec{e}_z}
{
\left[(\vec x'_\perp-\vec x_\perp)^2 + \tau^2 \sinh( \rap_0 \mp \eta)^2 \right]^{3/2}
},
\label{eq:magfieldspectators}
\end{multline}
where the proper time $\tau = (t^2 - z^2)^{1/2}$ and the space-time
rapidity $\eta = \frac{1}{2} \log[ (t + z)/(t-z)]$.

Many particles will be created by the interactions between the
participants. However, the number produced positively and negatively
charged particles will be nearly equal. Moreover the expansion of
these produced particles is almost spherical. We therefore expect that
the contribution of the produced particles to the magnetic field is
very small and will neglect this contribution. Hence we only have to
take into account the contribution of the participants which were
originally there. It is know that these participants stay
traveling along the beam axis according to the following normalized
distribution
\begin{equation}
  f(\rap) = \frac{a}{2 \sinh(a \rap_0)} e^{a \rap}, \;\;\;
  -\rap_0 \leq \rap \leq \rap_0.
\end{equation}
Experimental data shows that $a \approx 1/2$, consistent with the baryon junction stopping mechanism 
(see \cite{junction} and references therein).

The contribution of the participants to the magnetic field is hence
given by
\begin{multline}
 e \vec B^\pm_p(\tau, \eta, \vec x_\perp)  = 
\pm Z \alpha_{EM} \int \mathrm{d}^2 \vec x'_\perp
\int_{-\rap_0}^{\rap_0} \mathrm{d} \rap f(\rap)
\sinh(\rap \mp \eta)
\rho_\pm(\vec x'_\perp) \theta_\mp(\vec x'_\perp)  \times
\\
\frac{ (\vec x'_\perp - \vec x_\perp) \times \vec{e}_z}
{
\left[(\vec x'_\perp-\vec x_\perp)^2 + \tau^2 \sinh(\rap \mp \eta)^2 \right]^{3/2}
}.
\label{eq:bpart}
\end{multline}

\begin{figure}[htb]
\begin{center}
\includegraphics{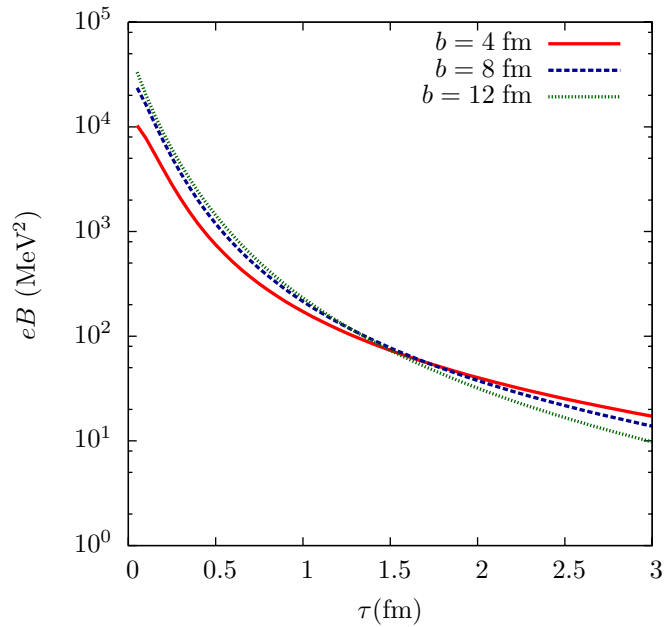}
\end{center}
\caption{Magnetic field at the center of a gold-gold
collision, for different impact
parameters. Here the center of mass energy is
$62\;\mathrm{GeV}$ per nucleon pair ($\rap_0 = 4.2$).}
\label{fig:magfield62}
\end{figure}

\begin{figure}[htb]
\begin{center}
\includegraphics{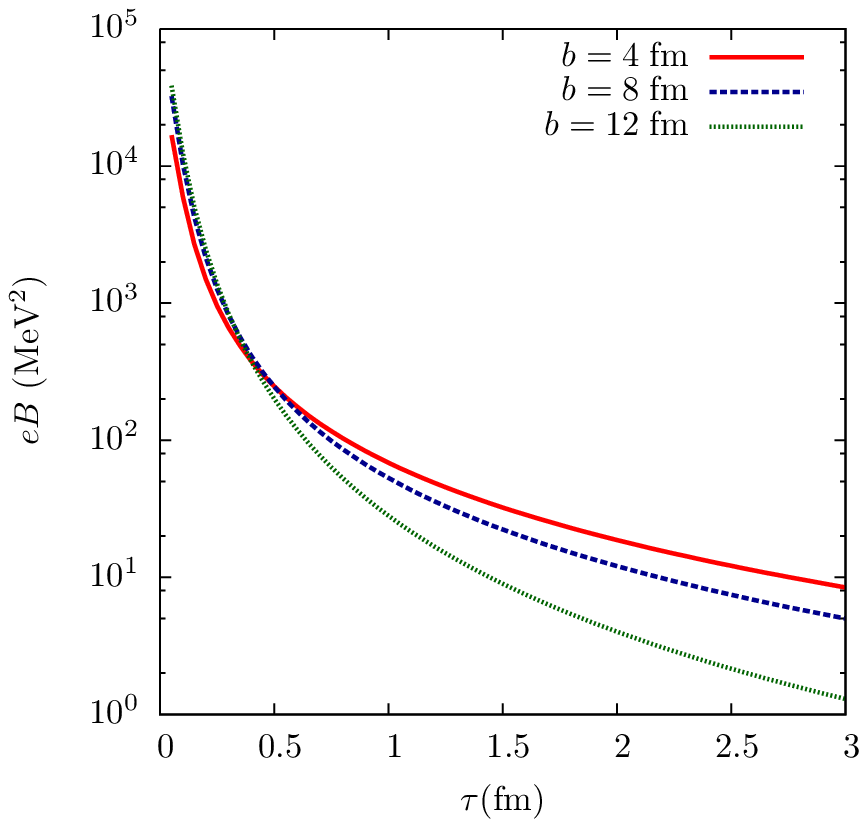}
\end{center}
\caption{Magnetic field at the center of a gold-gold
collision, for different impact
parameters.
Here the center of mass energy is
$200\;\mathrm{GeV}$ per nucleon pair ($\rap_0 = 5.4$).}
\label{fig:magfield200}
\end{figure}

We have evaluated the magnetic field numerically at the origin ($\eta
= 0, \vec x_\perp = 0$) in which case it is pointing in the
$y$-direction.  We took colliding gold ions ($Z=79$, $R =
7\;\mathrm{fm}$) with different beam rapidities ($Y=4.19$ and $Y=5.36$).
The results are displayed in Figs.~\ref{fig:magfield62} and
\ref{fig:magfield200}. Clearly enormous magnetic fields are created in
non-central heavy ion collisions.

For our purposes it is useful to have some analytic expression of the
magnetic field. We can find reasonable approximations for $\tau \gtrsim 2R /
\sinh(\rap_0)$. First we will consider the spectators, then we will
discuss an approximation for the participants.  We will perform both
approximations at the origin ($\vec x_\perp = 0$ and $\eta = 0$). In
that case the magnetic field is pointing in the $y$-direction, $e \vec
B = e B \vec e_y$.  Especially for large impact parameters the
magnetic field at the origin will be a good estimate for the magnetic
field at the surface of the interacting region, since the magnetic
field in the overlap region is to a good degree homogeneous in the
transverse plane.

\subsection{Spectator Contribution for $\tau \gtrsim R / \sinh(\rap_0)$}
For $\tau \gtrsim R \sinh(\rap_0)$ the denominator of
the integrand of the spectator contribution 
Eq.~(\ref{eq:magfieldspectators}) can
be approximated by $\tau^3 \sinh(\rap_0)^3$. Hence we find
\begin{equation}
 e B_s \approx Z \alpha_{EM} \exp(-2 \rap_0) \frac{4 R}{\tau^3} g(b/R),
\end{equation}
where
\begin{equation}
g(b/R) = \sum_\pm g_\pm(b/R),
 \label{eq:gimpactpardep}
\end{equation}
with
\begin{equation}
g_\pm(b/R) = \mp \frac{1}{R} \int \mathrm{d}^2 x_\perp'
\rho_\pm(\vec x_\perp')(1- \theta_\mp(\vec x_\perp')) x'.
\end{equation}
We find that to very good approximation $g_\pm(b/R) = b/R$. As a result
\begin{equation}
 e B_s \approx Z \alpha_{EM} \exp(-2 \rap_0) \frac{4 b}{\tau^3}.
 \label{eq:magfieldapproxspect}
\end{equation}

\subsection{Participant Contribution for $R/\sinh(\rap_0) \lesssim \tau \lesssim R$}
At midrapidity ($\eta = 0$) the participant contribution to the
magnetic field Eq.~(\ref{eq:bpart}) contains the following integral
over the beam rapidities
\begin{equation}
 I = \frac{a}{2 \sinh(a \rap_0)} \frac{1}{\tau^3}
 \int_{-\rap_0}^{\rap_0} \mathrm{d}\rap 
\frac{ \exp(a \rap) \sinh (\rap) }{[\beta^2 + \sinh^2 (\rap)]^{3/2}},
 \label{eq:ibeta}
\end{equation}
where $\beta^2 = (\vec x_\perp' - \vec x_\perp)^2 / \tau^2$.
We are interested in finding an approximation to Eq.~(\ref{eq:ibeta})
for $1 < \beta < \sinh(\rap_0)$. The first step is to write
\begin{equation}
 I \approx
\frac{a}{\exp(a \rap_0)} \frac{1}{2 \tau^3}
 \int_{0}^{\rap_0} \mathrm{d}\rap 
\frac{ \exp[ (1 + a) \rap] }{[\beta^2 + \tfrac{1}{4} \exp(2 \rap)]^{3/2}}.
\end{equation}
The last integral can be computed exactly. However, since
it is quite complicated for $a=1/2$ and because we are
anyway interested in the limit where $1 < \beta < \sinh(\rap_0)$
we will expand the result in $\beta / \sinh(\rap_0)$.
This gives for $a < 1$,
\begin{equation}
 I \approx c \exp(-a Y_0)
\frac{1}{\tau^{1+a}} \frac{1}{ \vert \vec x_\perp' - \vec x_\perp \vert^{2-a}},
\end{equation}
where
\begin{equation}
 c =  \frac{2^{1+a} a \Gamma(1-a/2) \Gamma(3/2 + a/2)}{(1+a)\sqrt{\pi}}.
\end{equation}
For $a=1/2$ we find $c=0.59907\ldots$.

\begin{figure}[htb]
\begin{center}
\includegraphics{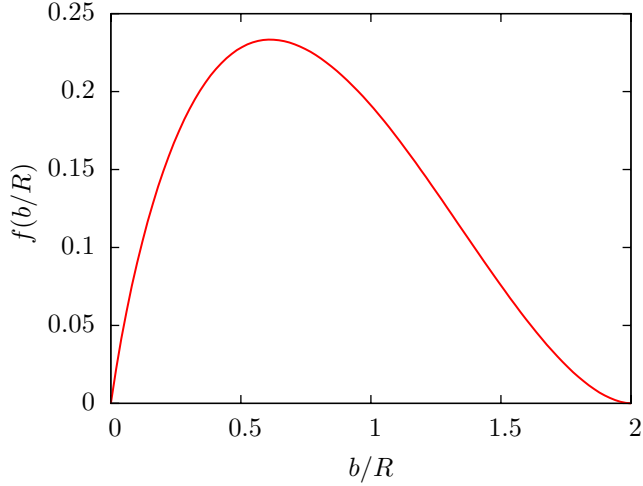}
\end{center}
\label{fig:f}
\caption{Plot of the function $f(b/R)$. This function
describes the impact parameter dependence of the
approximated participant contribution to the magnetic
field and is defined in Eq.~(\ref{eq:impactpardep}).}
\end{figure}

Now let us try to approximate the magnetic field at the origin $(\eta
= 0, \vec x_\perp =0)$ using the approximation for $I$. 
Taking $a=1/2$ we find that
\begin{equation}
 e B_p \approx c Z \alpha_{EM} \exp(-Y_0/2) \frac{1}{R^{1/2} \tau^{3/2}} 
f(b/R),
 \label{eq:magfieldapprox}
\end{equation}
where
\begin{equation}
f(b/R) = \sum_\pm f_\pm(b/R),
 \label{eq:impactpardep}
\end{equation}
with
\begin{equation}
f_\pm(b/R) = \mp R^{1/2} \int \mathrm{d}^2 x_\perp'
\rho_\pm(\vec x_\perp')\theta_\mp(\vec x_\perp') \frac{x'}{\vert \vec x'_\perp \vert^{3/2}}.
\end{equation}
We have computed $f(b/R)$ numerically and displayed the result in
Fig.~(\ref{fig:f}).

\section{Alternative derivation of Chiral Magnetic Effect}
\label{app:veccurrent}
We will show in a somewhat more formal way than in
Sec.~\ref{sec:mechanism} that a homogeneous background magnetic
field in the presence of a gauge field configuration with nonzero
$\qw$ gives rise to a vector current $j_\mu = \langle \bar \psi
\gamma_\mu \psi \rangle_A$ along the direction of the magnetic field.

We take the background magnetic field to point in the $z$-direction so
that only the following component is non-vanishing,
\begin{equation}
 j_3 = \langle \psi \gamma_3 \psi \rangle_A = 
 \langle \psi_R^\dagger \sigma_3 \psi_R \rangle_A
-
\langle \psi_L^\dagger \sigma_3 \psi_L \rangle_A.
 \label{eq:j3} 
\end{equation}
Initially, there is no difference between the number of left-handed
and right-handed handed modes, therefore $j_3$ vanishes and there is
no charge separation. Clearly Eq.~(\ref{eq:j3}) shows that if left-
and right-handed modes are polarized in the same direction (as it
should in a large magnetic field), a charged current will only form if
there is a difference between $N_R$ and $N_L$. More explicitly in the
limit of a very large magnetic field ($eB \gg T^2$ or $eB \gg p_F^2$, where
$p_F$ is the Fermi momentum)
\begin{equation}
\int \mathrm{d}^3 x\, 
\langle \psi_{R,L}^\dagger \sigma_3 \psi_{R,L}\rangle_A = 
\mathrm{sgn}(q) 
\int \mathrm{d}^3 x\, 
\langle \psi^\dagger_{R,L} \psi_{R,L}\rangle_A
.
\end{equation}
Here $q$ denotes the charge of the particle and reflects the fact that
particles with different charge are polarized oppositely in a magnetic
field.  We now immediately find that in the limit of large magnetic
field the total current for a flavor $f$ is given by
\begin{equation}
  J^f_3 = \int \mathrm{d}^3x j^f_3(x) = \mathrm{sgn}(q_f) (N^f_R - N^f_L).
\end{equation} 

In the chiral limit this implies that at $t=\infty$ 
an electromagnetic (EM) and a baryon (B) current
is formed which magnitude equals
\begin{equation}
  J^{EM}_3 = 2 \qw \sum_f \vert q_f \vert, 
  \;\;\;\;\;\;\;
  J^{B}_3 = \tfrac{2}{3} \qw \sum_f \mathrm{sgn}(q_f).
\end{equation}

If we assume that all quarks were located within a finite volume,
this electromagnetic current generates a charge difference
$Q$ between two sides of a plane perpendicular to the magnetic
field of size
\begin{equation}
 Q = 2 \qw \sum_f \vert q_f \vert.
\end{equation}
This difference is the same as we found before in Eq.~(\ref{eq:deltaq}).

In the same way we find the following baryon number difference 
\begin{equation}
 B = 2 \sum_f J^f_3 = \tfrac{2}{3} Q_w \sum_f \mathrm{sgn}(q_f).
\end{equation}
The baryon number difference vanishes for $N_f = 2$, while
for $N_f = 3$, $B=\tfrac{2}{3} \qw$. 
Since $B$ times the average charge of a baryon is smaller than $Q$, we
expect the charge difference mainly to be realized in mesons, like
$\pi^+$ and $\pi^-$.

In this appendix we considered a gauge field configuration with
nonzero $\qw$ in the presence of a very large homogeneous magnetic
field. It would be interesting to extend these results by studying the
behavior of the vector current in a smaller magnetic field.

\end{document}